\PassOptionsToPackage{table,xcdraw}{xcolor}

\documentclass[10pt,conference]{IEEEtran}

\usepackage{alltt                                    % I like these
	, multirow
	,subfig
	, booktabs
	, listings
	, graphicx
	,float
	,cite
	,verbatim
	,mathtools
	,url
}

\usepackage{xcolor}
\usepackage{xcolor, soul}
\usepackage[numbers]{natbib}
\usepackage{threeparttable}
\usepackage{framed}
\usepackage{expl3}
\usepackage{hyperref}
\usepackage{filecontents}

\usepackage{pgfplots}
\pgfplotsset{compat=1.16}
\usepackage{pgfplotstable}

\usepgfplotslibrary{statistics}

\usepackage{scalefnt}
\usepackage{tikz}

\usepackage{tabularx}
\usepackage{lipsum}
\usepackage{array}

\usepackage{tcolorbox}
\usepackage{multirow}

\usetikzlibrary{shapes.gates.logic.US,trees,positioning,arrows}
\usetikzlibrary{shapes.multipart}
\usetikzlibrary{shapes,arrows}

\tikzstyle{vertex}=[ellipse,fill=black!25,minimum size=20pt, inner sep=0pt]
\tikzstyle{edge} = [draw,thin,-]
\tikzstyle{glabel} = [text width=1cm,text centered,font=\bf]
\pgfdeclarelayer{bg}    % declare background layer
\pgfsetlayers{bg,main} 

\ExplSyntaxOn
\newcommand\latinabbrev[1]{
  \peek_meaning:NTF . {% Same as \@ifnextchar
    #1\@}%
  { \peek_catcode:NTF a {% Check whether next char has same catcode as \'a, i.e., is a letter
      #1., \@ }%
    {#1., \@}}}
\ExplSyntaxOff

\usepackage{tikz}
\usepackage{verbatim}
\usetikzlibrary{arrows,shapes,backgrounds}

\tikzstyle{vertex}=[ellipse,fill=black!25,minimum size=20pt, inner sep=0pt]
\tikzstyle{edge} = [draw,thin,-]
\tikzstyle{glabel} = [text width=1cm,text centered,font=\bf]
\pgfdeclarelayer{bg}    % declare background layer
\pgfsetlayers{bg,main}  

\usepackage{soul}
\usepackage{pifont}
\usepackage{booktabs,makecell}
\usepackage[export]{adjustbox}
\usepackage{balance}
\usepackage{paralist}

\usepackage{pgf-pie}

\newif\ifpienumberinlegend
\pgfkeys{/number in legend/.code=
    \expandafter\let\expandafter\ifpienumberinlegend
    \csname if#1\endcsname
    \ifpienumberinlegend

    \def\beforenumber##1\afternumber{}%
    \fi,
    /number in legend/.default=true
}

\makeatletter
\pgfplotsset{
    boxplot/hide outliers/.code={
        \def\pgfplotsplothandlerboxplot@outlier{}%
    }
}
\makeatother

\newcommand{\CASE}[1]{\STATE \textbf{case} #1\textbf{:} \begin{ALC@g}}
\newcommand{\ENDCASE}{\end{ALC@g}}

\newcommand{\DEFAULT}{\STATE \textbf{default:} \begin{ALC@g}}
\newcommand{\ENDDEFAULT}{\end{ALC@g}}
\newcommand{\DEFAULTLINE}[1]{\STATE \textbf{default:} }
\newsavebox{\supbox}% Superscript box
\newcommand{\bsup}{\begin{lrbox}{\supbox}$\tt\scriptstyle}% Superscript begin
\newcommand{\esup}{$\end{lrbox}{}^{\usebox{\supbox}}}% Superscript end

\definecolor{lightpurple}{rgb}{0.8,0.8,1}
\definecolor{codebg}{RGB}{255,255,255}
\definecolor{commentcolor}{RGB}{11,140,11}
\usepackage[normalem]{ulem} % for \sout
\usepackage{xcolor}

\usepackage{ifthen}
\usepackage{amssymb}
\newboolean{showcomments}
\setboolean{showcomments}{false}
\ifthenelse{\boolean{showcomments}}
{\newcommand{\nbc}[3]{
 {\colorbox{#3}{\bfseries\sffamily\scriptsize\textcolor{white}{#1}}}
 {\textcolor{#3}{\sf\small$\blacktriangleright$\textit{#2}$\blacktriangleleft$}}
 }
}
{\newcommand{\nbc}[3]{}
 }

\usepackage{amsmath}
\usepackage{enumitem}

\newcommand*\rectangled[4]{%
  \tikz[baseline=(char.base)]{%
    \node[shape=rectangle, fill=#2, draw=#3, text=#4, inner sep=1pt] (char) {#1};%
  }%
}

\begin{document}

% \title{Predicting Unanswered Questions of Stack Overflow During their Submission}
% \title{Early Detection of Unanswered Questions and Evidence-based Guideline to Increase Their Answerability}

% \title{Does Your Question at Stack Overflow Need a Code Snippet? Investigating the Cause \& Effect of Missing Code Snippets}
% \title{Can We Identify Stack Overflow Questions Requiring Code Snippets? Investigating the Cause \& Effect of Missing Code Snippets}
\title{GENCNIPPET: Automated Generation of Code Snippets for Supporting Programming Questions}

\author{
\IEEEauthorblockN{Saikat Mondal}
\IEEEauthorblockA{University of Saskatchewan\\
\ saikat.mondal@usask.ca}
\and
% \IEEEauthorblockN{Mohammad Masudur Rahman}
% \IEEEauthorblockA{Dalhousie University\\
% \ masud.rahman@dal.ca}
% \and
\IEEEauthorblockN{Chanchal K. Roy}
\IEEEauthorblockA{University of Saskatchewan\\
\ chanchal.roy@usask.ca}
}
% \author{\IEEEauthorblockN{Mohammad Masudur Rahman}
% \IEEEauthorblockA{Dalhousie University\\
% \ masud.rahman@dal.ca}
% }
% \author{\IEEEauthorblockN{Chanchal K. Roy}
% \IEEEauthorblockA{University of Saskatchewan\\
% \ chanchal.roy@usask.ca}
% }

% \author{Anonymous Authors}

% \author{Saikat Mondal}
% \affiliation{%
%   \institution{University of Saskatchewan, Canada}
% %   \city{Saskatoon}
% %   \country{Canada}
%     }
% \email{saikat.mondal@usask.ca}

% \author{Banani Roy}
% \affiliation{%
%   \institution{University of Saskatchewan, Canada}
% %   \city{Saskatoon}
% %   \country{Canada}
%   }
% \email{banani.roy@usask.ca}

% \affiliation{%

%PredUnations

% \author{\IEEEauthorblockN {Saikat Mondal\hspace{0.7cm}Avijit Bhattacharjee\hspace{0.7cm}Chanchal K. Roy}
% \IEEEauthorblockA{SRlab, Department of Computer Science, University of Saskatchewan\\ }}
% \ saikat.mondal@usask.ca}
% }

% \renewcommand{\shortauthors}{Saikat Mondal et al.}
% \renewcommand{\shorttitle}{Reproducibility Challenges and Their Impacts on Technical Q\&A Websites}

\maketitle

\begin{abstract}
\textit{Context:} Software developers often ask questions on Technical Q\&A forums like Stack Overflow (SO) to seek solutions to their programming-related problems (e.g., errors and unexpected behavior of code).

\smallskip
\textit{Problem:} Many questions miss required code snippets due to the lack of readily available code, time constraints, employer restrictions, confidentiality concerns, or uncertainty about what code to share. Unfortunately, missing but required code snippets prevent questions from getting prompt and appropriate solutions.

\smallskip
\textit{Objective:} We plan to introduce \texttt{GENCNIPPET}, a tool designed to integrate with SO's question submission system. \texttt{GENCNIPPET} will generate relevant code examples (when required) to support questions for their timely solutions.

\smallskip
\textit{Methodology:} We first downloaded the SO April 2024 data dump, which contains 1.94 million questions related to Python that have code snippets and 1.43 million questions related to Java. Then, we filter these questions to identify those that genuinely require code snippets using a state-of-the-art machine learning model. 
Next, we select questions with positive scores to ensure high-quality data. 
% We exclude questions containing multiple code snippets and finalize our dataset. 
Our plan is to fine-tune Llama-3 models (e.g., Llama-3-8B), using 80\% of the selected questions for training and 10\% for validation. The primary reasons for choosing Llama models are their open-source accessibility and robust fine-tuning capabilities, which are essential for deploying a freely accessible tool.
\texttt{GENCNIPPET} will be integrated with the SO question submission system as a browser plugin. It will communicate with the fine-tuned model to generate code snippets tailored to the target questions. The effectiveness of the generated code examples will be assessed using automatic evaluation against ground truth, user perspectives, and live (wild) testing in real-world scenarios.

\end{abstract}


%________________________
\section{Introduction}
\label{sec:introduction}
%________________________

The advent of technical Q\&A forums like Stack Overflow (SO) has significantly transformed the landscape of programming problem-solving \cite{vasilescu2014social}. Technical forums revolutionize developer collaboration by offering global expertise, collective problem-solving, diverse solutions, searchable knowledge, community validation, reputation building, and continuous learning \cite{barua2014developers, ponzanelli2014mining, sillito2008asking}. Stack Overflow remains relevant despite the increasing use of large language models (LLMs) for coding assistance. Our analysis of Python and Java question trends from 2022 to early 2025 using the Stack Exchange API reveals that SO  received 434,997 new Python-related questions, while Java-related had 139,516. While overall SO activity has declined since the release of LLMs like ChatGPT, specific programming communities remain active with no significant drop in question frequency \cite{da2024chatgpt}. Replacing SO entirely with LLMs is impractical, as over-reliance on AI reduces learning opportunities, limits solution diversity, and hinders critical thinking \cite{lee2025impact, liu2023better}. Instead, integrating LLMs to enhance SO questions can create an optimal balance between automated efficiency and human expertise.

\begin{figure}[!htb]
	\centering
	\includegraphics[width = 3.2in]{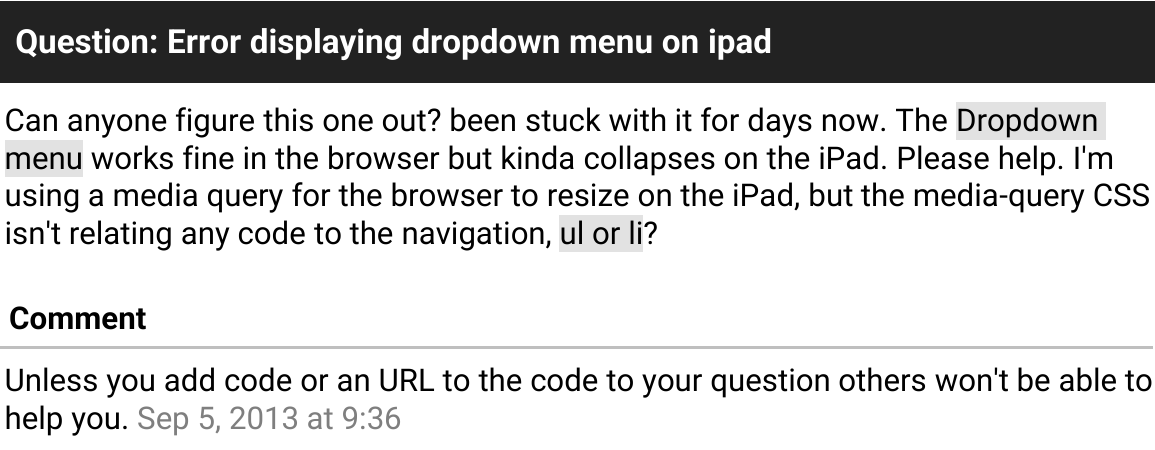}
	\caption{Motivating examples}
	\label{fig:motivationExamples}
	% \vspace{-5mm}
\end{figure}

A large number of SO questions discuss the programming problems (e.g., coding errors, unexpected behavior) that warrant code snippets for a resolution \cite{treude2011programmers, squire2014bit}. 
% Traditionally, users at SO first analyze the code snippets to identify or reproduce the reported problems \cite{emse2021bmondal}. Upon success, they can submit appropriate solutions.
Unfortunately, question submitters often miss the required code snippets, which prevents these questions from getting appropriate answers promptly \cite{mondal2024can, calefato2018ask, duijn2015quality, asaduzzaman2013answering, treude2011programmers}. 
For example, the question shown in Fig. \ref{fig:motivationExamples} discusses an issue related to the display of a drop-down menu on an iPad. However, the question submitter did not include the problematic code snippet in the question. As a result, users who attempted to answer the question could not figure out the problem. One user thus commented, 
``{\fontfamily{lmss}\selectfont Unless you add code or a URL to the code to your question, others will not be able to help you}.'' 
Unfortunately, the question submitter still did not add any code snippets, and the question did not receive any answers over the years.

A recent study by Mondal et al. \cite{mondal2024can} shows that only 23.8\% of questions get acceptable answers that miss the required code snippets. On the contrary, such a percentage is 61.4\% for the questions that include code snippets during their submission. The delay in getting acceptable answers is also significantly higher for the questions that miss the code during submission \cite{mondal2024can}. Such a scenario might explain the 31\%  unanswered and more than 50\% unresolved questions at SO \cite{datadumpapi,rahman2015insight,asaduzzaman2013answering}.
Unanswered questions and those lacking acceptable solutions pose direct and indirect challenges to developers. As an example of a direct challenge, developers struggle to find answers to their programming issues that might lead to development delays or adopt sub-optimal solutions. Indirectly, questions that lack high-quality solutions can contaminate training data for code search/generation tools (e.g., ChatGPT). It can impact their ability to learn accurate patterns, reducing their quality and effectiveness. As a result, developers will be affected globally, resulting in reduced productivity and low-quality work.

\begin{figure*}[t]
	\centering
	\includegraphics[width = 7in]{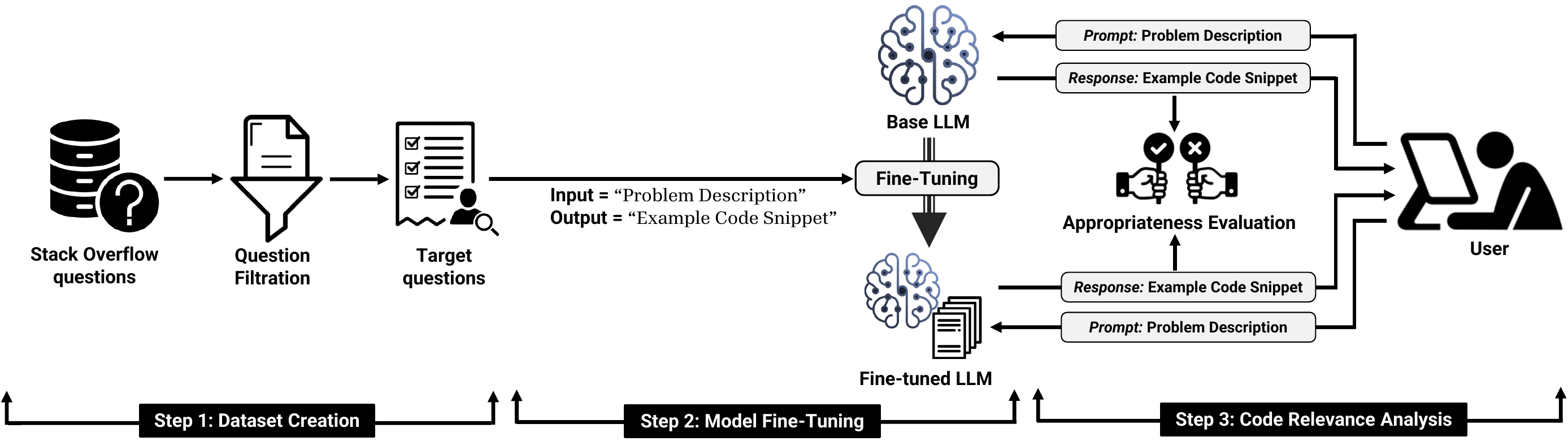}
	\caption{Study methodology}
	\label{fig:studyMethodology}
	% \vspace{-4mm}
\end{figure*}

Mondal et al. \cite{mondal2024can} initially propose machine learning techniques to identify questions that miss code snippets. However, simply identifying the questions missing code snippets may not be sufficient, as 29.7\% of users do not include code snippets because they lack readily available code, 45.3\% face employer restrictions, 45.3\% worry about sharing confidential information, 42.2\% are short on time, and 51.6\% are unsure which parts of the code to include \cite{mondal2024can}. In this study, we thus plan to introduce \texttt{GENCNIPPET}, a tool powered by LLM, to generate example code snippets to support SO questions for their timely resolution. It balances AI capabilities and human expertise to preserve the collaborative, learning-oriented nature of technical Q\&A like SO.

Unlike tools that generate complete coding solutions, \texttt{GENCNIPPET} focuses on generating example snippets, avoiding AI over-reliance, and mitigating the risks of incorrect implementations. While newer LLMs have improved, research highlights their limitations in generating reliable programming solutions. Studies show that LLM-generated solutions are error-prone, with API misuse rates as high as 62\% and incorrect implementations in 52\% of cases \cite{kabir2024stack, zhong2024can}. Misuse of APIs can lead to resource leaks and crashes \cite{zhong2024can}. Additionally, 39\% of users fail to detect misinformation in ChatGPT's responses, and developers use only 5.83\% of ChatGPT-generated code as-is due to reliability concerns \cite{das2024investigating, das2025why, kabir2024stack}. While LLMs provide speed, they struggle with debugging-related queries, making them less reliable for practical problem-solving \cite{oishwee2024large}. These findings reinforce the need for peer-reviewed, expert-assessed content on platforms like SO. By enhancing question quality with example snippets, \texttt{GENCNIPPET} strengthens SO as a robust, accessible programming knowledge base, ensuring long-term utility and reliability for the global developer community.

% This study aims to generate example code snippets to support programming questions using an LLM-powered tool instead of generating solutions. Generating complete solutions with LLMs leads to over-reliance on AI and risks introducing critical errors. Existing studies suggest that coding solutions generated by LLMs are often prone to errors like API misuse (62\%) and incorrect implementations (52\%) \cite{kabir2024stack, zhong2024can}. Developers engage with LLM tools like ChatGPT primarily for ideation and use only 5.83\% of ChatGPT-generated code as-is to resolve their issues due to lack of reliance \cite{das2024investigating, das2025why}.
% Additionally, while LLMs like ChatGPT are valued for their speed, they have knowledge cut-offs. They could not replace the accuracy and contextual reliability provided by SO's peer-reviewed, expert-assessed content. By enhancing question quality through example code snippets, we contribute to the growth of a robust, freely accessible, invaluable programming knowledge base of SO, ensuring long-term reliability and practical utility for the developer community.

% -----------------------------
\section{Research Question} 
\label{sec:research-questions}
% -----------------------------

This study aims to support SO questions that miss required code snippets by providing relevant example code. We formulate three key research questions as follows.

\begin{description}

    \smallskip
    \item[RQ1:] \textbf{How well do foundation models perform in generating relevant code examples for questions that miss required code snippets?}
    This research question addresses a critical challenge in technical Q\&A forums like SO—improving the clarity and completeness of questions that lack required code snippets. Foundation models, such as GPT-4, provide general-purpose code generation but may lack precision in aligning with SO-style snippet generation. Evaluating their effectiveness will provide insights into their limitations and help establish the need for fine-tuning.
    
    % \smallskip
    % \noindent \textbf{We define the following null hypothesis (H1):} A fine-tuned LLM model cannot generate contextually relevant example code snippets to support the quality of questions.

    \smallskip
    \item[RQ2:] \textbf{How does a fine-tuned LLM model compare to foundation models in generating relevant code snippets for SO questions?}
    This research question highlights the trade-off between specialization and generalization in AI systems. With their broad training objectives, foundation models (e.g., GPT-4) perform diverse tasks but may not generate code snippets that precisely reflect the intent of SO question-askers. A fine-tuned LLM model tailored for SO-specific data aims to bridge this gap by improving contextual alignment and relevancy. We can determine the benefits of domain-specific adaptation by comparing performance across models.

    % \smallskip
    % \noindent \textbf{We define the following null hypothesis (H2):} A fine-tuned LLM model does not outperform foundation models in generating contextually relevant code snippets to enhance question quality.

    \smallskip
    \item[RQ3:] \textbf{How useful is GENCNIPPET in assisting users with generating code snippets for their Stack Overflow questions?}
    The impact of our fine-tuned LLM model can be measured by its ability to assist SO users during question submission by generating relevant code examples for those who lack ready-to-use code or face restrictions in including it. We thus propose an online tool, GENCNIPPET, designed to provide real-time support. The tool aims to enhance question quality by offering required code snippets, thereby improving the chances of receiving timely and appropriate solutions. Its usability and effectiveness will be evaluated through the real-world usage of actual SO users.
    
    % \smallskip
    % \noindent \textbf{We define the following null hypothesis (H3):} GENCNIPPET does not help users generate relevant code examples.
    
\end{description}

% \usepackage{booktabs}
\begin{table}[htb]
    \centering
    \caption{Dataset construction summary}
    \label{table:dataset-construction}
    \resizebox{3.4in}{!}{%
    \begin{tabular}{l|c|c|c}
    \toprule
    & \textbf{\begin{tabular}[c]{@{}c@{}}Questions with\\ Code Snippets\end{tabular}} & \textbf{\begin{tabular}[c]{@{}c@{}}Questions Needing\\ Code Snippets\end{tabular}} & \textbf{\begin{tabular}[c]{@{}c@{}}Questions with Positive Score\\  and Single Code Snippet\end{tabular}} \\ \midrule
    
        \textbf{Java}   & 1,432,458  & 1,295,213  & 242,494      \\ \midrule
        \textbf{Python} & 1,939,742  & 1,781,793   & 316,058     \\ \midrule
        \textbf{Total}  & 3,372,200  & 3,077,006   & 558,552     \\ \bottomrule

    \end{tabular}
    }
\end{table}

%----------------------------------------------------------------------------------
\section{Generation of Example Code Snippets for SO Questions Using Foundation and Fine-Tuned LLMs (RQ1 \& RQ2)}
\label{sec:example-code-generation}
%----------------------------------------------------------------------------------

% \section{Study Methodology}
% \label{sec:methodology}

Fig. \ref{fig:studyMethodology} shows the methodology of generating example code snippets using a fine-tuned LLM model and evaluating their relevance. The following sections discuss different steps in detail.

\subsection{Dataset Construction}
\label{subsec-mothodology:createDataset}

Fig. \ref{fig:studyMethodology} (Step 1) shows the steps of our dataset construction, while Table \ref{table:dataset-construction} summarizes the dataset. We first downloaded the SO April 2024 data dump—the most recent dataset available at the start of this study—from the Stack Exchange site \cite{datadumpapi}. To ensure our findings remain relevant to current development trends, we will incorporate more recent data up to the point when we begin fine-tuning.

The dataset contains 1.94 million questions related to Python that have code snippets and 1.43 million questions related to Java. However, questions do not always benefit from the inclusion of code examples. In some cases, code snippets are optional or even detrimental to question quality \cite{chua2015answers, baltadzhieva2015predicting, mondal2024can}. To address this, we identify questions that genuinely require code snippets by leveraging the state-of-the-art machine learning model proposed by Mondal et al. \cite{mondal2024can}.

To further ensure the quality and relevance of code snippets, we select questions with positive scores. To reduce noise, we initially excluded questions containing multiple code snippets. As a result, our dataset consists of 558,552 questions (242,494 in Java and 316,058 in Python), each with a positive score and one code snippet. However, we plan to include a subset of questions with multiple code snippets to assess their impact on model performance. 
% Additionally, we incorporate the time factor in our dataset to ensure API relevance and prevent outdated information.

\subsection{Model Fine-Tuning}
\label{subsec-mothodology:model-fine-tuning}

The foundational LLMs are general-purpose models and often lack the necessary contextual focus. As a result, they may struggle to generate code examples that are relevant and appropriate to the specific context of a question. However, existing studies suggest that fine-tuned models with high-quality task-specific data help models better understand the context to perform specific tasks \cite{jin2023inferfix, li2024exploring, li2020dlfix, jiang2021cure, nashaat2024towards, yu2024fine}.
Therefore, we plan to fine-tune LLM base models, specifically the open-source Llama-3 (e.g., Llama-3-8B), due to its accessibility and cost-effectiveness. We will randomly select 80\% of Java- and Python-related questions from our dataset for fine-tuning, with 10\% used for validation and 10\% for testing. We plan to use LoRA (Low-Rank Adaptation), a parameter-efficient fine-tuning approach that significantly reduces memory requirements to ensure practical feasibility. For example, when fine-tuning Llama-3-8B, we may require a GPU with at least 16 GB VRAM (e.g., RTX 3090) for basic training, while 24 GB VRAM (e.g., RTX 4090) provides better flexibility and performance.
If the fine-tuned Llama model demonstrates satisfactory performance, we will not consider any paid models (e.g., GPT-4). However, if the results indicate limitations, we will explore alternative models such as GPT-4 and CodeLlama for comparison. We plan to conduct instructional fine-tuning by providing structured input-output pairs to enhance alignment with SO-style question structures.

To fine-tune the LLM model, we will follow these steps:

\begin{itemize} 

    \item\textbf{Input/Output Data Preparation:} We will extract problem descriptions and their corresponding code snippets from SO questions. Programming language specifications will be extracted from question tags.
    
    \item \textbf{Input Formatting:} We structure inputs by combining the problem description, programming language tag, and a timestamp to provide temporal context and clear guidance to the model (e.g., ``Question: [text] Language: [Java] Date: [2023-06-01]''). This helps the model align its responses with the relevant API version and development practices at the time of the post.
    
    %\textbf{Input Formatting:} We structure inputs by combining the problem description and programming language tag (e.g., ``Question: [text] Language: [Java]'') to provide clear guidance to the model.
    
    \item \textbf{Output Formatting:} We structure outputs as code snippets in the format: ``Code: [code]''
    
    \item \textbf{Model Initialization:} We plan to load the pre-trained Llama-3 (e.g., Llama-3-8B) model, which is optimized for natural language and code generation tasks.
    
    \item \textbf{Fine-Tuning Configuration:} We will configure training parameters, including learning rate (2e-5), batch size (32), and the number of epochs (3, with early stopping). We also plan to use a cross-entropy loss function to minimize discrepancies between generated and expected outputs.
    
    \item \textbf{Training Process:} We will fine-tune the model using the prepared input-output pairs while monitoring validation loss throughout the training process to ensure effective learning and mitigate overfitting. 
    
    % \item \textbf{Model Evaluation:} The trained model will be evaluated using ROUGE, BLEU, and BERTScore for automatic assessment. Additionally, we will conduct a manual evaluation on 400 randomly selected samples to assess relevance and correctness.

\end{itemize}

While our current approach plans to fine-tune LLMs on static data, we also plan to explore RAG technology or API search integration to enhance snippet accuracy and adaptability to evolving frameworks.

\subsection{Code Relevance Evaluation}
\label{subsec-mothodology:code-relevance}

We plan to evaluate the relevance and correctness of the generated code snippets using the following three ways.

\smallskip \textbf{Automatic Evaluation:} We will assess all test samples automatically using similarity-based metrics to measure syntactic and semantic closeness between generated and ground-truth code snippets. The evaluation will employ ROUGE, BLEU, and BERTScore, which provide different perspectives on code similarity. Since these metrics do not always capture relevance, automatic evaluation will serve as an initial filtering step, complemented by manual review.

\smallskip \textbf{Manual Review:} Recognizing that automatic evaluation may not fully capture generated code's clarity and relevance \cite{hu2022correlating}, we will manually evaluate 400 randomly selected samples (95\% confidence level, 5\% margin of error). Two experts specializing in Java and Python will assess the clarity and relevance of the snippets using a 5-point Likert scale. This evaluation will help identify cases where generated snippets deviate from the question's intent.
In addition to evaluating clarity and relevance, this manual assessment will help identify cases where the generated snippets introduce unintended issues not present in the original question. Such issues can affect the reproducibility of reported problems and reduce the practical reliability of the tool. Quantifying these errors will offer critical insights for improving generation quality and ensuring the tool's effectiveness in real-world development scenarios.

\smallskip \textbf{Wild Testing:} To validate the real-world effectiveness of generated snippets, we will conduct a wild test on SO using the latest questions posted within the last three months. We will submit AI-generated code snippets as suggested edits to 50 selected questions that originally lacked them. Since SO requires that all edits undergo community review, we will track the acceptance rate of our suggestions as an indicator of the practical utility of our approach. Additionally, we will analyze whether questions with added snippets receive faster responses or improved engagement. This method ensures that the generated code examples align with real developer expectations and contribute meaningfully to the SO knowledge base.

\subsection{Performance Comparison with Foundation Models}
\label{subsec-mothodology:performance-comparison}

We compare fine-tuned models with foundation models (e.g., GPT-4) for generating relevant code snippets. We use standardized prompt engineering (zero-shot and few-shot prompting) for foundation models to ensure fairness. The same set of questions from Section \ref{subsec-mothodology:code-relevance} will be used.
Evaluation includes automatic (ROUGE, BLEU, BERTScore) and manual assessments. To isolate the impact of fine-tuning, we compare: 
\begin{itemize} 
    \item Fine-tuned Llama-3 (e.g., Llama-3-8B) 
    \item Prompt-tuned GPT-4 (zero-shot and few-shot prompting) 
    \item Zero-shot GPT-4 
    \item Code-specific LLM (e.g., CodeLlama) 
\end{itemize}

This analysis determines whether fine-tuning improves snippet relevance beyond general-purpose models and examines failure cases to refine our approach.

The example prompt designed for generating relevant code snippets is as follows.

\noindent\rule{\linewidth}{1pt}

\noindent {\fontfamily{lmss}\selectfont\small I am working on a programming problem and need help generating a representative code example to demonstrate my programming issue.}

\smallskip
\noindent {\fontfamily{lmss}\selectfont\small \textbf{[Problem Description]:} The detailed description of the problem discussed in questions, including context (the situation where the issue occurs, such as a specific library, framework, or environment), expected behavior (explain what the code is supposed to do), and the specific issue (describe the type of fault the code should have, such as logical errors, syntax issues, or unexpected runtime behavior).}

\smallskip
\noindent {\fontfamily{lmss}\selectfont\small \textbf{[Programming Language]:} Python, Java}

\smallskip
\noindent {\fontfamily{lmss}\selectfont\small \textbf{[Constraints and Requirements]:} Any specific constraints, such as a required library, specific input/output, or code style.}

\smallskip
\noindent {\fontfamily{lmss}\selectfont\small \textbf{[Objective]:} Generate a code example that aligns with the provided details, demonstrates the problem clearly, and contains intentional faults related to the described issue. Keep the example concise and focused to ensure clarity for troubleshooting.}

\noindent\rule{\linewidth}{1pt}

\begin{figure}[htb]
	\centering
	%\resizebox{4.5in}{!}{%
	\includegraphics[width=3in]{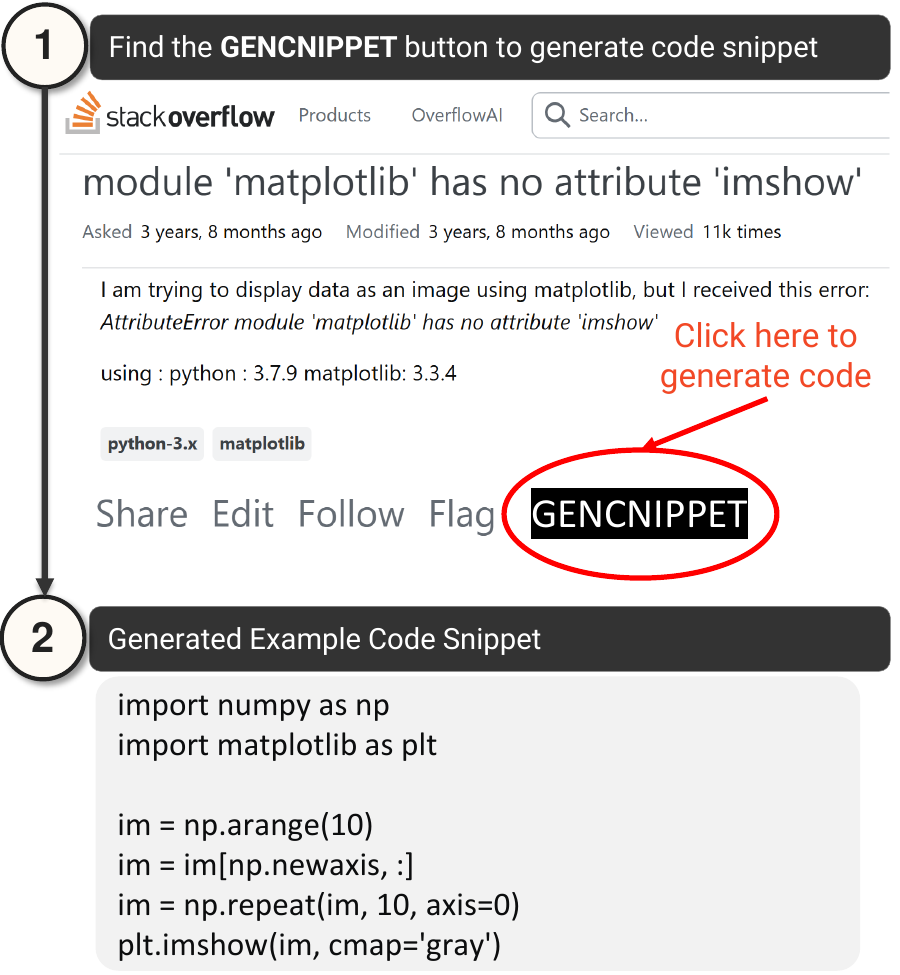}
	%}
	\caption{Proposed GENCNIPPET interface.}
	\label{fig:GENCNIPPET-interface}
\end{figure}

% ---------------------------------------------------------------
\section{GENCNIPPET: A Tool for Generating Example Code Snippets (RQ3)}
\label{sec:tool-support}
% ---------------------------------------------------------------

We propose GENCNIPPET, a web-based browser plug-in designed to enhance SO question submission by generating example code snippets. Powered by a fine-tuned LLM, GENCNIPPET analyzes the problem descriptions in question texts and generates relevant code snippets to improve question clarity and quality.
To assess its practical utility, we plan to conduct a real-world study involving SO users. This study will provide valuable insights into its usability, effectiveness, and limitations in enhancing the question submission process.

\begin{figure}[htb]
	\centering
	%\resizebox{4.5in}{!}{%
	\includegraphics[width=3.4in]{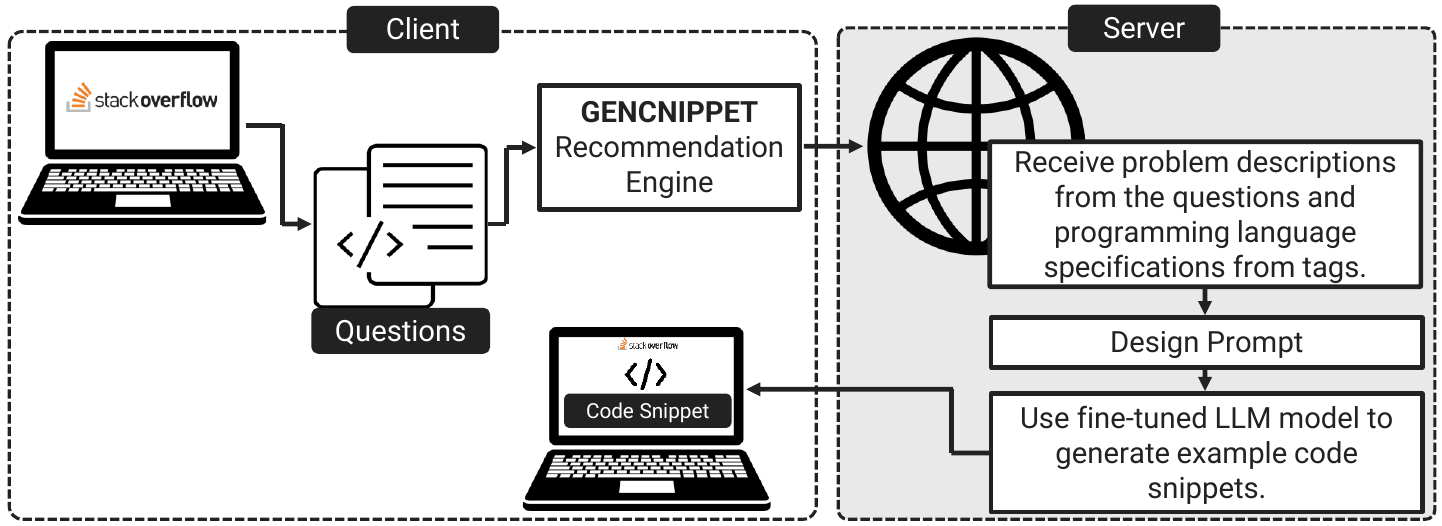}
	%}
	\caption{An overview of the GENCNIPPET system architecture.}
	\label{fig:GENCNIPPET-system-architecture}
\end{figure}

\subsection{GENCNIPPET Interface and Architecture}
\label{subsec:GENCNIPPET-architecture}

Fig. \ref{fig:GENCNIPPET-interface} shows the proposed interface of GENCNIPPET. Once GENCNIPPET is installed as a browser plug-in, a dedicated \rectangled{GENCNIPPET}{black}{black}{white}
% {\fontfamily{lmss}\selectfont\small ``GENCNIPPET''} 
button (Fig. \ref{fig:GENCNIPPET-interface}(1)) will appear within the question submission interface of SO. Users can generate example code snippets to support their questions by clicking this button.
%\circled{1}{black}{white}{white}

Fig. \ref{fig:GENCNIPPET-system-architecture} illustrates the GENCNIPPET architecture, which consists of two main components: the client and the server.
On the client side, users interact with the GENCNIPPET interface, which includes a button for generating code snippets. When users click this button, the client-side script captures the problem description and programming language specifications from the question. This data is then sent to the server-side application.
On the server side, the application constructs a prompt using the received problem description and programming language details. This prompt is submitted to the fine-tuned LLM model to generate a relevant example code snippet that supports the problem description. Once the code snippet is generated, it is sent back to the client-side script.
Finally, the client-side script presents the generated code snippet to the user, ready to be included in their question. This workflow ensures seamless integration and enhances the quality of user-submitted questions with contextually relevant code examples.

% ---------------------------------------------------------------
\subsection{Effectiveness and Utility Evaluation Plan of GENCNIPPET}
\label{subsec:tool-effectiveness}
% ---------------------------------------------------------------

We plan to recruit at least twenty SO users to evaluate GENCNIPPET. Participants will be asked to install the browser plug-in and use it to generate example code snippets while submitting programming-related questions. Each participant will be requested to submit at least three questions using GENCNIPPET-generated snippets. After completing the tasks, participants will provide feedback through a survey, which will be analyzed to assess the tool's effectiveness and user experience.

\subsection{Survey}
\label{subsec:survey}

\noindent\textbf{Survey Participants.} We plan to recruit active users of SO who are currently working with either Java or Python programming languages using the following two ways.

\begin{itemize}
    \item \emph{Snowball Approach:} We will use convenience sampling to bootstrap the snowball \cite{stratton2021population}. First, we will contact a few software developers who are active users of SO and are known to us, easily reachable, and working in software companies worldwide. We will explain our study goals and share the online survey with them. We will then adopt a snowballing method \cite{bi2021accessibility} to disseminate the survey to several of their colleagues with similar experiences.

    \smallskip
    \item \emph{Open Circular:} We will also circulate the survey to specialized Facebook groups. In particular, we will target the groups where professional software developers discuss their programming problems. We also plan to use LinkedIn to find potential participants because it is one of the largest professional networks.
\end{itemize}

\noindent After finalizing the participants, we will provide a video tutorial with step-by-step instructions on installing and using GENCNIPPET. Following the completion of their assigned tasks using the tool, we will collect their responses to evaluate their experience and gather feedback.

\smallskip
\noindent\textbf{Survey Design.} Our survey comprises the following sections.

\begin{itemize}
    \item \emph{Consent and Prerequisite:} This section ensures participants' informed consent to participate in the survey and their agreement to the processing of their data. It also specifies the prerequisites for participation, such as participants must be active SO users and currently working with either Java or Python programming languages.

    \smallskip
    \item \emph{Participants' Information:} In this section, we gather relevant demographic and professional details, including participants' programming experience, gender, current profession, organization, SO profile information, and country. This information helps contextualize the survey findings and analyze variations in user experiences.

    \smallskip
    \item \emph{Ease of Use of GENCNIPPET:} This section aims to evaluate the usability and user experience of the GENCNIPPET tool. We will ask participants to rate the following statements based on their experience.

    \begin{description}
        \item[1.] How easy was it to install and set up GENCNIPPET on your browser? (Very easy/Easy/Neutral/Difficult/Very difficult) 

        \item[2.] Did you encounter any technical issues while using GENCNIPPET? (No issues/Minor issues/Moderate issues/Major issues)

        \item[3.] How would you rate the response time (latency) of GENCNIPPET in generating code snippets? (Very fast/Fast/Acceptable/Slow/Very slow)

        \item[4.] Please provide any additional comments or suggestions for improving the ease of use of GENCNIPPET (Open-ended question)
        
    \end{description}

    \smallskip
    \item \emph{Net Promoter Score (NPS):} This section measures how likely participants recommend GENCNIPPET to others, helping us assess user satisfaction.

    \begin{description}
        \item[1.] Would you recommend GENCNIPPET to other SO users? (Definitely/Probably/Neutral/Probably not/Definitely not)

        The Weighted Net Promoter Score (WNPS) is calculated as: 

        \begin{equation}
            WNPS = \frac{\sum (S_i \times N_i)}{N_{\text{total}}}
        \end{equation}
        where:
        $S_i$  is the score assigned to each response type (2, 1, 0, -1, -2),
        $N_i$ is the number of respondents selecting that response,
        $N_{\text{total}}$ is the total number of responses.

        \item[2.] What is the primary reason for the score you gave? (Open-ended question)

        \item[3.] What improvements would make you more likely to recommend GENCNIPPET to others? (Open-ended question)
        
    \end{description}

    \smallskip
    \item \emph{Utility Rating of GENCNIPPET:} This section evaluates the practical benefits and effectiveness of GENCNIPPET in enhancing the question submission process on SO. We plan to ask participants to rate the following statements based on their experience with the tool.

    \begin{description}
        \item[1.] The code snippet generation feature saved your time during question submission (Strongly agree/Agree/Neutral/Disagree/Strongly disagree)

        \item[2.] The generated code snippets were relevant to the problem descriptions in your questions (Strongly agree/Agree/Neutral/Disagree/Strongly disagree)
        
        \item[3.] The generated code snippets were clear, concise, and appropriately tailored to the intent of my questions (Strongly agree/Agree/Neutral/Disagree/Strongly disagree)
        
        \item[4.] Using GENCNIPPET made you feel more confident about receiving timely answers to my questions (Strongly agree/Agree/Neutral/Disagree/Strongly disagree)
        
        \item[5.] GENCNIPPET reduced the effort required to create high-quality questions(Strongly agree/Agree/Neutral/Disagree/Strongly disagree)
        
        \item[6.] GENCNIPPET improved my overall experience of submitting questions on SO (Strongly agree/Agree/Neutral/Disagree/Strongly disagree)
        
        \item[7.] Please share any additional thoughts or suggestions about the utility of GENCNIPPET (Open-ended question)
    
    \end{description}

\end{itemize}

\smallskip
\noindent\textbf{Survey Response Analysis Plan.}
The survey responses will be analyzed to evaluate the usability, satisfaction, and utility of GENCNIPPET. Participant data will first be validated to ensure prerequisites are met, followed by a descriptive analysis of demographic information to contextualize findings. Usability will be assessed through frequency distributions and thematic analysis of ease-of-use responses. Net Promoter Score (NPS) will be calculated to measure satisfaction, with open-ended feedback analyzed for improvement suggestions. Utility ratings will be summarized using descriptive statistics and cross-tabulated with demographic data to identify trends. Statistical tests will explore significant patterns, and open-ended responses will provide qualitative insights. Findings will be reported with visualizations and actionable recommendations to improve GENCNIPPET’s functionality and user experience.

% ---------------------------------------------------------------
\section{Implications of the Study}
\label{sec:study-implications}
% ---------------------------------------------------------------

GENCNIPPET will enhance SO questions by providing missing but required code snippets. Questions with problem descriptions and required code snippets receive faster and more appropriate solutions \cite{treude2011programmers, mondal2024can, calefato2018ask}. By automatically generating relevant snippets, GENCNIPPET helps in framing high-quality questions \cite{asaduzzaman2013answering, squire2014bit}.

It also addresses challenges users face in sharing code due to confidentiality concerns or uncertainty about what to include \cite{mondal2024can}. This can lead to quicker responses and improved community engagement \cite{calefato2018ask, duijn2015quality}. Enhancing question quality with code snippets may reduce unanswered and unresolved questions \cite{rahman2015insight, asaduzzaman2013answering}, reinforcing SO as a valuable programming knowledge base. Its integration as a browser plug-in ensures seamless adoption without disruption to workflow.

User feedback and real-world testing will refine GENCNIPPET to meet diverse needs. Its adoption can streamline question submission, save time, and boost confidence in receiving timely solutions. Over time, this could elevate question standards, reduce moderation workload, and foster a more supportive knowledge-sharing environment. 
% Additionally, by enhancing question quality, GENCNIPPET can improve AI-driven tools that rely on SO for training.

% ---------------------------------------------------------------
\section{Threat to Validity}
\label{sec:threat-to-validity}
% ---------------------------------------------------------------

\textbf{External Validity.}
Our study focuses on SO questions related to Java and Python, covering both statically and dynamically typed languages. This suggests that GENCNIPPET may work for other similar programming languages. However, generalizing to languages with fundamentally different paradigms, such as functional or domain-specific languages, remains uncertain. While SO remains a widely used platform, the evolution of developer behavior with emerging AI tools may impact the long-term applicability of our findings. Additionally, our user study involves a limited number of SO users, and their experience may not fully represent the broader SO community.

\textbf{Internal Validity.}
The effectiveness of GENCNIPPET depends on the quality of the dataset and fine-tuning process. While we ensure dataset relevance by filtering for questions requiring code snippets, biases in data selection or errors in code generation could impact the model’s performance. Moreover, user behavior in real-world testing may vary—users might accept AI-generated snippets without thorough validation, affecting the interpretation of usability results.

\textbf{Construct Validity.}
The evaluation of generated code snippets relies on automated metrics (ROUGE, BLEU, BERTScore) and manual expert review. While these metrics capture text similarity, they may not fully reflect clarity and relevance. To mitigate this, we conduct manual assessments. However, subjective biases in human evaluation and variations in how developers perceive snippet relevance may introduce inconsistencies. To address this, we involve two reviewers and use a structured evaluation framework to assess the clarity and relevance of generated code.

\textbf{Survey Validity.}
Our user study follows a snowball sampling approach, which may introduce participant bias. We thus plan to recruit participants through an open circular. Additionally, self-reported feedback in surveys is prone to bias—participants may provide overly positive or negative responses based on their expectations rather than actual tool performance. We will ensure anonymity in responses to encourage honest feedback and cross-analyze survey responses.

\section*{Acknowledgment}
This research is supported in part by the Natural Sciences and Engineering Research Council of Canada (NSERC) Discovery Grants program, the Canada Foundation for Innovation's John R. Evans Leaders Fund (CFI-JELF), and by the industry-stream NSERC CREATE in Software Analytics Research (SOAR).

\balance

% \begin{acks}
% This research is supported by the Natural Sciences and Engineering Research Council of Canada (NSERC), and by a Canada First Research Excellence Fund (CFREF) grant coordinated by the Global Institute for Food Security (GIFS).
% \end{acks}

% \balance
% \bibliographystyle{plainnat}
\bibliographystyle{unsrtnat}
\bibliography{reference}

\end{document}